\documentstyle[11pt,newpasp,twoside,epsf]{article}
\def\edcomment#1{\iffalse\marginpar{\raggedright\sl#1\/}\else\relax\fi}
\reversemarginpar

\begin{document}
\title{The Fate of Luminous Compact Blue Galaxies: An Environmental Approach}

\author{S. M. Crawford, M. A. Bershady, A. D. Glenn, and J. G Hoessel}
\affil{Department of Astronomy, University of Wisconsin--Madison, 
475 N. Charter St, Madison, WI 57315}

\begin{abstract}

Luminous Compact Blue Galaxies (LCBGs) are a heterogeneous class which
dominate an intermediate phase of galaxy evolution.  These sources
account for the majority of the star formation between $0.3<z<1$, yet
the identity of their present-day counterparts is an open question.
An environmental dependence to their evolution may provide the answer.
We have undertaken the first census of LCBGs in intermediate-redshift
clusters using broad and narrow-band images from the WIYN Long Term
Variability Study.  Several key clusters are in the southern sky.  The
Southern African Large Telescope's (SALT) Prime Focus Imaging
Spectrograph (PFIS), with a field of view matching our WIYN data, will
allow us to determine several fundamental characteristics of cluster
LCBGs, including (1) star formation rates (SFR) and metallicities from
low dispersion {\it m}ulit-{\it o}bject {\it s}pectroscopy (MOS); (2)
dynamical masses from line-widths measured via high dispersion MOS;
and (3) cluster velocity dispersions of LCBGs relative to other
cluster emission-line galaxies via Fabry-Perot imaging or MOS. We
tentatively connect these distant galaxies with their local, evolved
counterparts, and discuss how PFIS observations will enable us to make
the physical distinction between LCBGs in clusters and the field.

\end{abstract}

\section{Introduction}

A key prediction of CDM hierarchical formation models (e.g., White \&
Frenk 1991) is the ``down-sizing'' of star formation sites, whereby
the characteristic galaxy mass dominating the co-moving star-formation
rate decreases with redshift (Cowie et al. 1996). At high redshift,
Lyman-break galaxies dominate; we believe they are associated with
today's high-mass systems (Lowenthal et al. 1997; Steidel et
al. 1996). Today, the most intensely star-forming systems are low-mass
HII galaxies. Between $0.1 < z < 1.0$ the situation is less clear.

``Luminous compact blue galaxies'' (LCBGs) appear to be the most
rapidly evolving class of galaxy at intermediate redshift (Lilly et
al. 1998; Mallen-Ornelas et al. 1999). These enigmatic galaxies,
initially identified by Koo \& Kron (1988), are luminous ($M_B \sim
-20$), small ($R_e \sim 2-3$kpc), and massive engines of star
formation (up to $\sim40 M_{\odot} yr^{-1}$) (Koo et al. 1995,
Guzm\'an et al. 1998, Hammer et al. 2000).  They appear to form a link
in redshift, size, and luminosity between Lyman-break galaxies and HII
galaxies today (Lowenthal et al. 1997; see our Figure 1).  LCBGs are a
major contributor to the observed enhancement of the star-formation
density of the universe at $z\la1$, and their mass and number density
decline in concert with the rapid drop in the global SFR since $z=1$
(Guzm\'an et al. 1998).

\begin{figure}
\plotfiddle{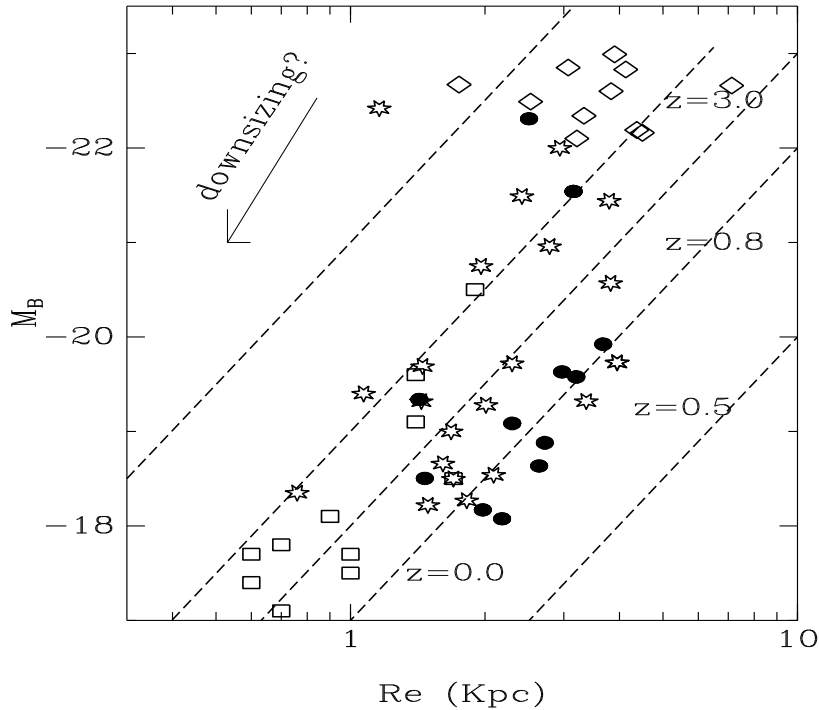}{3.25in}{0}{65.}{45.}{-215}{-75.}
\vskip 0.25in
\caption{Luminosity and half-light radius for different types of
compact star-bursting galaxies: a subsample of local star-bursts from
the UCM survey (open squares; Pisano et al. 2001); intermediate
redshift LCBGs (stars; Phillips et al. 1998); and Lyman Break galaxies
(diamonds, Lowenthal et al. 1997). Filled circles are field and
cluster LCBGs from the cluster-field MS0451 at $z\sim0.5$ with
magnitudes and radii measured from a WFPC2 F702 image. 
The arrow indicates the qualitative
effects of ``down-sizing,'' nearly along (dashed) lines of constant
rest-frame surface brightness.}
\end{figure}

\begin{figure}
\plotfiddle{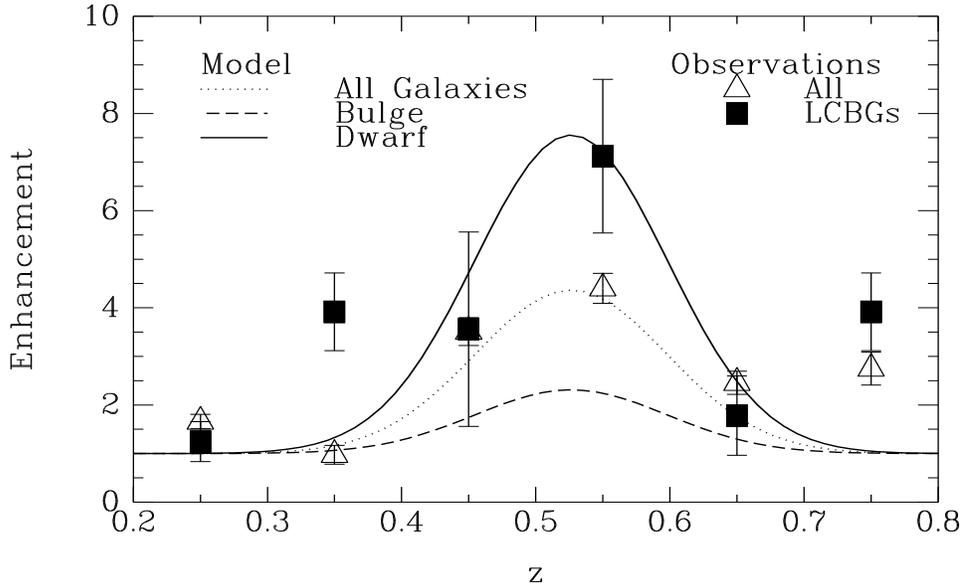}{2.5in}{270}{55.}{55.}{-225.}{245.}
\vskip 0.15in
\caption{Galaxy enhancement as a function of photometric redshift for
MS0451. The dotted line predicts the enhancement of a Coma-like
cluster for all galaxy types. The dashed and solid lines predict the
LCBG enhancements if they are predecessors of spiral bulges or
spheroidals, respectively. Triangles and squares are the {\it
observed} enhancement of all galaxies and LCBGs, respectively --
consistent with LCBGs being associated to a bust-phase of cluster dwarfs.}
\vskip -0.15in
\end{figure}

However, a debate continues about today's descendants of LCBGs. Due to
their narrow emission line-widths and number density, Koo et
al. (1995) proposed LCBGs are the progenitors of today's lower-mass
spheroidal (Sphs) galaxies: If their current burst of star formation
terminates, LCBGs fade by 4 mag in a few Gyrs, yielding luminosities,
surface-brightnesses, sizes, and colors similar to, e.g., NGC 205
(Guzm\'an et al. 1998).  LCBGs have also been proposed as the
period when bulges of spiral galaxies form due to their high gas phase
metallicities (Hammer et al. 1999, Kobulnicky \& Zaritsky 1999). In
this picture, the bright, blue compact regions are embedded within a
lower-surface brightness, disk of larger size, (Barton \& Van Zee
2001).

\section{LCBGs in Galaxy Clusters:  New Clues on LCBGs Descendants}

Because of the extreme density of galaxy clusters, they offer a
distinctly different, yet heretofore unexplored environment in which
to study the nature of LCBGs. For our purpose, there are two salient
differences in galaxy populations in rich clusters and the field:
First, the morphology-density relationship describes a decreasing
fraction of spirals with increasing local surface density of galaxies
(Dressler 1980). Second, the dwarf-galaxy population is much richer in
clusters. Dwarf galaxies are the most numerous type of galaxy
regardless of environment, but the ratio of dwarf to giant galaxies
increases with the overall density (e.g., Trentham and Hodgkins 2002).
Because the relative densities of the two proposed descendants of
LCBGs (lower-mass spheroidals and spiral bulges) are different in
galaxy clusters and the field, the number density of LCBGs in
intermediate galaxy clusters make the connection between the past
(LCBGs) and today (either low-mass spheroidals or
spirals). Specifically, if LCBGs are relatively more numerous in
intermediate-redshift clusters than in the field, this provides
independent evidence that LCBGs are associated with today's lower-mass
spheroidal population.

We currently are analyzing imaging data on 10 intermediate redshift
galaxy clusters from the WIYN Long Term Variability Survey (WLTV) to
search for cluster LCBGs. Each cluster has been repeatedly imaged over
5 years in a 10 arcmin field with the WIYN 3.5m telescope. We have a
total of $2-5$ hrs integration per target in each of the UBRIz bands
taken under excellent seeing conditions (typically $<$ 0.75''). Deep,
archival HST images are available of each cluster core.  We are in the
process of gathering rest-frame OII and continuum narrow-band images
of the 6 deepest clusters, four of which are visible to SALT.

We have completed the first stage of an LCBG search in cluster MS0451,
at $z=0.53$. We frame our analysis in the measurement of galaxy
enhancement. The enhancement is the number of galaxies at a given
redshift divided by the expected number in the field at that redshift
(estimated from comparable deep images without rich clusters). Hence,
if there is no galaxy over-density, the enhancement is one. Figure 2
shows the enhancement of LCBGs in the field of MS0451. This is a
robust, {\it differential} measurement, which, for this cluster
appears to support the hypothesis that LCBGs evolve into Sphs.

\begin{figure}
\plotfiddle{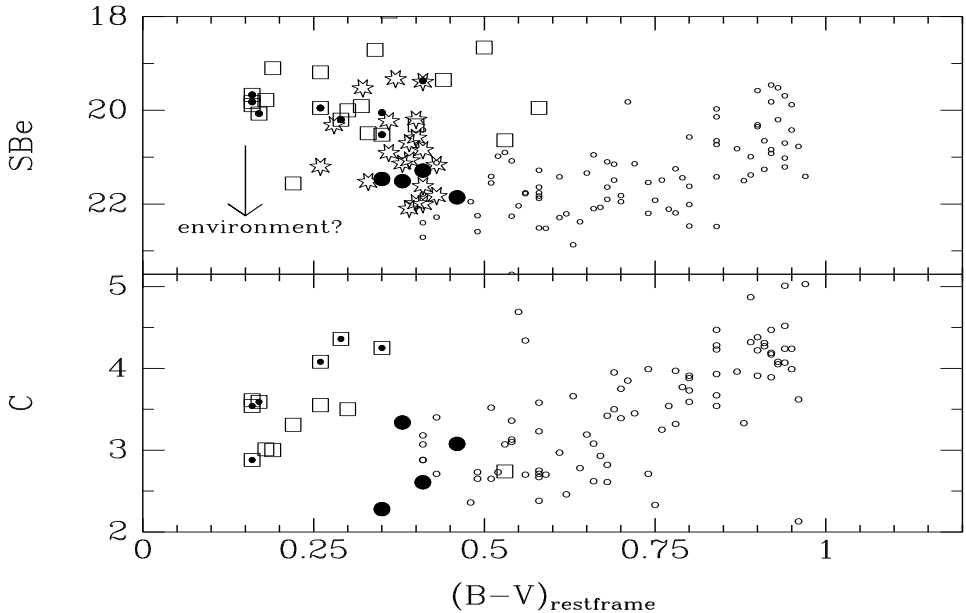}{2.75in}{270}{60}{50}{-250}{250}
\vskip 0.15in
\caption{Rest-frame surface-brightness (top) and image concentration
(bottom) versus color for a selection normal, nearby galaxies (small
points; Bershady et al. 2000) and LCBGs: Compact Narrow Emission Line
Galaxies (squares) and Blue Nucleated Galaxies (squares with dots;
Guzm\'an et al. 1997, 1998); field LCBGs (stars; Phillips et al. 1998);
and LCBGs appearing in the core of MS0451 (filled circles). Cluster
LCBGs have similar color and concentration, but are lower in 
surface-brightness than field LCBGs.}
\end{figure}

We also have calculated the photometric and structural characteristics
of the LCBGs appearing in the inner 0.75 Mpc of MS0451 using F702
WFPC2 images. Cluster LCBGs also have similar colors, sizes, and
profiles (based on image concentration) to field LCBGs, but somewhat
lower luminosities or surface-brightnesses (see Figures 1 and 3). Are
these differences indicative of a more rapid pace of ``down-sizing''
in clusters, akin to the environmental dependence of star formation
rates (Balogh et al. 1998, Martin et al. 2000)? To differentiate
between an accelerated downsizing scenario and environmental effects
of truncated or quenched star-formation (due, e.g., to stripping),
star formation rates {\it and} mass measurements of cluster LCBGs are
required.  These measurements, well suited to PFIS, will tell us what
differs between cluster and field: the mass-function of star-bursts, or
the star-formation rate at a given mass?

\section{LCBG Properties:  The Future with PFIS}
 
Preliminary results from our survey raise several important questions
about cluster LCBG requiring spectroscopic measurements on 10m-class
telescopes to answer. SALT's Prime Focus Imaging Spectrograph (PFIS)
will enable the needed high throughput, medium-resolution
spectroscopy. Here we outline how we plan to use PFIS for our LCBG
studies.

\begin{itemize}

\item {\bf Mass Uniformity:} Our initial evidence indicates
photometric differences in cluster and field LCBGs. We can
characterize this difference more meaningfully by comparing field to
cluster LCBG masses. In making this measurement for several clusters,
we will map how the mass of star-bursting galaxies evolves with
redshift and environment. Because of their small sizes, LCBG virial
masses are measured by combining spatially-unresolved emission
line-widths and HST or WIYN size measurements. The highest spectral
resolutions of PFIS are ample for LCBG line-width measurements. By
using PFIS's bank of narrow band ($\frac{\lambda}{\Delta \lambda} \sim
50$) filters, we can limit the range of the dispersed spectra on the
detector to $\sim$75\AA\ (rest-frame) around the [OII]$\lambda$3727
emission line for each cluster. This allows us to increase the spatial
multiplexing of our multi-object masks (see the MMS concept described
by Bershady et al., these proceedings) -- and hence also our survey
efficiency -- by about a factor of 4. The minimum source spatial
separation in the dispersion direction is $\sim$1 arcmin at the
highest spectral resolutions (grating angle $\alpha=50^\circ$), and
decreases linearly with resolution ($\propto sin \alpha / cos \alpha$,
i.e., the Littrow condition for the VPH gratings).  At the resolution
used by Guzman et al. (1998), the minimum separation is 10 arcseconds.

\item {\bf Star Formation Rates:} If field and cluster LCBGs have
similar size and mass functions, luminosity or surface-brightness
differences are caused by lower star-formation rates -- plausibly due
to gas stripping in the dense clusters environments. We can measure
these rates via low-resolution PFIS MOS observations that capture
[OII]$\lambda$3727 and other blue nebular lines that will allow us
also to determine metallicity and estimate extinction. Combined mass,
size, shape, color, star-formation characteristics will allow us to
carefully quantify physical differences between field and cluster
LCBGs, and determine if they share similar histories of stellar
processing of baryonic matter.

\item {\bf Origins: Cluster Accretion?} There is ample evidence in low
redshift clusters that blue, star-forming galaxies are on the
periphery of these systems, and just falling in. One way to test for
an accretion origin of intermediate-redshift cluster LCBGs is to
compare their velocity dispersion to the red cluster population (for
which some measurements already exist, and will be augmented). LCBG
velocity dispersions can be calculated efficiently from the MOS
measurements described above, or via Fabry-Perot (FP) scanning with
the lowest-resolution PFIS etalon. The advantage of the latter is the
gain in spatial multiplex over the full 8 arcmin PFIS field, thereby
providing complete kinematic characterization of the emission-line
population in the cluster out to several cluster radii.

\end {itemize}

\section{Summary}

We have presented the initial results from the WLTV survey that will
provide an extensive catalog of $\sim200$ cluster and field LCBGs with
high quality photometric information in areas around 10 clusters
between $0.3<z<1$. With cluster LCBGs and over-densities identified
from these photometric data, we plan to use PFIS to answer deeper
questions about LCBGs: Are cluster and field LCBGs the same? Does the
evolution of cluster and field LCBGs differ? What is the origin of
cluster LCBGs? The answers to these questions tie in with parallel
studies of other cluster and field populations, and will yield a
cohesive picture of galaxy evolution over the past 3-7 Gyr.

\acknowledgments This program has been funded by NSF AST-9970780 and
AST-0307417, and NASA LTSA NAG5-6032 and STScI/AR-9917. Graduate
student travel to this conference was funded by the NSF.


\begin{references}

\reference Balogh, M.~L., et al. 1998, ApjL, 504, L75
\reference Barton, E.~J.~\& van Zee, L.\ 2001, ApJl, 550, L35
\reference Bershady, M.~A., Jangren, A., \& Conselice, C.~J.\ 2000, AJ , 119, 2645
\reference Cowie, L. L., et al.\ 1996 AJ, 112, 839
\reference Dressler, A.\ 1980, ApJ , 236, 351
\reference Guzm\'an, R., et al.\ 1997, ApJ , 489, 559
\reference Guzm\'an, R., et al.\ 1998, ApJl, 495, L13
\reference Hammer, F., et al.\ 2001, ApJ , 550, 570
\reference Kobulnicky, H.~A.~\& Zaritsky, D.\ 1999, ApJ , 511, 118
\reference Koo, D.~C.~\& Kron, R.~G.\ 1988, ApJ , 325, 92 
\reference Koo, D.~C., et al.\ 1995, ApJl, 440, L49
\reference Lilly, S.~et al.\ 1998, ApJ, 500, 75 
\reference Lowenthal, J.~D.~et al.\ 1997, ApJ , 481, 673
\reference Mall{\' e}n-Ornelas, et al.\ 1999, ApJl, 518, L83
\reference Martin, C.~L., Lotz, J., \& Ferguson, H.~C.\ 2000, ApJ , 543, 97 
\reference Phillips, A.~C., et al. 1997, ApJ , 489, 543
\reference Pisano, D.~J., et al. \ 2001, AJ, 122, 1194
\reference Steidel, C.~C., et al. \ 1996, AJ, 112, 352 
\reference Trentham N. \& Hudgkin, S. \ 2002, astro-ph 0202437
\reference White, S.~D.~M.~\& Frenk, C.~S.\ 1991, ApJ, 379, 52 

\end{references}
\end{document}